\documentclass{stacs_proc}

\usepackage[latin1]{inputenc}
\usepackage{amssymb}
\usepackage{amsmath}
\usepackage{dsfont}
\usepackage{xspace}
\usepackage{array}

\newcommand{\progress}[1]{\stackrel{#1}{\longrightarrow}}

\newcommand{\Progress}[1]{\stackrel{#1}{\Longrightarrow}}

\begin{document}

\title{On Termination for Faulty Channel Machines}

\author[lab1]{P. Bouyer}{Patricia Bouyer}

\author[lab1]{N. Markey}{Nicolas Markey}

\author[lab2]{J. Ouaknine}{Jo\"el Ouaknine}

\author[lab1]{Ph.\ Schnoebelen}{Philippe Schnoebelen}

\author[lab2]{J. Worrell}{James Worrell}

\address[lab1]{LSV, ENS Cachan, CNRS\newline
61 Av.\ Pdt.\ Wilson, F-94230 Cachan, France\newline
\texttt{\{bouyer,markey,phs\}@lsv.ens-cachan.fr}}

\address[lab2]{Oxford University Computing Laboratory\newline 
Wolfson Bldg., Parks Road, Oxford OX1 3QD, UK\newline
\texttt{\{joel,jbw\}@comlab.ox.ac.uk}}

\thanks{Thanks: Patricia Bouyer is also affiliated with the Oxford University
  Computing Laboratory and is partially supported by a Marie Curie
  Fellowship.}


\begin{abstract}
\noindent 
A \emph{channel machine} consists of a finite controller together with
several fifo channels; the controller can read messages from the head
of a channel and write messages to the tail of a channel. In this
paper, we focus on channel machines with \emph{insertion errors},
i.e., machines in whose channels messages can spontaneously
appear. Such devices have been previously introduced in the study of
Metric Temporal Logic. We consider the \emph{termination} problem: are
all the computations of a given insertion channel machine finite? We
show that this problem has non-elementary, yet primitive recursive
complexity.
\end{abstract}

\maketitle

\stacsheading{2008}{121-132}{Bordeaux}
\firstpageno{121}

\bibliographystyle{plain}

\section{Introduction}

Many of the recent developments in the area of automated verification,
both theoretical and practical, have focussed on infinite-state
systems. Although such systems are not, in general, amenable to fully
algorithmic analysis, a number of important classes of models with
decidable problems have been identified. Several of these classes,
such as Petri nets, process algebras, process rewrite systems, faulty
channel machines, timed automata, and many more, are instances of
\emph{well-structured transition systems}, for which various problems
are decidable---see~\cite{FS01} for a comprehensive survey.

Well-structured transition systems are predicated on the existence of
`compatible well-quasi orders', which guarantee, for example, that
certain fixed-point computations will terminate, etc. Unfortunately,
these properties are often non-constructive in nature, so that
although convergence is guaranteed, the rate of convergence is not
necessarily known. As a result, the computational complexity of
problems involving well-structured transition systems often remains
open.

In this paper, we are interested in a particular kind of
well-structured transition systems, known as faulty channel
machines. A channel machine (also known as a queue automaton) consists
of a finite-state controller equipped with several unbounded fifo
channels (queues, buffers). Transitions of the machine can write
messages (letters) to the tail of a channel and read messages from the
head of a channel. Channel machines can be used, for example, to model
distributed protocols that communicate asynchronously.

Channel machines, unfortunately, are easily seen to be Turing
powerful~\cite{BZ83}, and all non-trivial verification problems
concerning them are therefore undecidable.
In~\cite{AJ93,finkel94,CFP96,AJ96}, Abdulla and Jonsson, and Finkel
\emph{et al.}\ independently introduced \emph{lossy channel machines}
as channel machines operating over an unreliable medium; more
precisely, they made the assumption that messages held in channels
could at any point vanish nondeterministically.  Not only was this a
compelling modelling assumption, more adequately enabling the
representation of fault-tolerant protocols, for example, but it also
endowed the underlying transition systems of lossy channel machines
with a well-structure, thanks to Higman's lemma~\cite{Higman52}. As a
result, several non-trivial problems, such as control-state
reachability, are decidable for lossy channel machines.

Abdulla and Jonsson admitted in~\cite{AJ93} that they were unable to
determine the complexity of the various problems they had shown to be
decidable. Such questions remained open for almost a decade, despite
considerable research interest in the subject from the scientific
community. Finally, Schnoebelen showed in~\cite{schnoebelen02} that
virtually all non-trivial decidable problems concerning lossy channel
machines have non-primitive recursive complexity. This result, in
turn, settled the complexity of a host of other problems, usually via
reduction from reachability for lossy channel machines. Recently, the
relevance of the lossy channel model was further understood when it
was linked to a surprisingly complex variant of Post's correspondence
problem~\cite{CS07}.

Other models of unreliable media in the context of channel machines
have also been studied in the literature. In~\cite{CFP96}, for
example, the effects of various combinations of insertion,
duplication, and lossiness errors are systematically
examined. Although insertion errors are well-motivated (as former
users of modems over telephone lines can attest!), they were
surprisingly found in~\cite{CFP96} to be theoretically uninteresting:
channels become redundant, since read- and write-transitions are
continuously enabled (the former because of potential insertion
errors, the latter by assumption, as channels are
unbounded). Consequently, most verification problems trivially reduce
to questions on finite automata.

Recently, however, slightly more powerful models of channel machines
with insertion errors have appeared as key tools in the study of
Metric Temporal Logic (MTL)\@. In~\cite{OW05,OW06a}, the authors
showed that MTL formulas can capture the computations of insertion
channel machines \emph{equipped with primitive operations for testing
channel emptiness}. This new class of faulty channel machines was in
turn shown to have a non-primitive recursive reachability problem and
an undecidable recurrent control-state reachability
problem. Consequently, MTL satisfiability and model checking were
established to be non-primitive recursive over finite
words~\cite{OW05}, and undecidable over infinite words~\cite{OW06a}.

Independently of Metric Temporal Logic, the notion of emptiness
testing, broadly construed, is a rather old and natural one. Counter
machines, for instance, are usually assumed to incorporate primitive
zero-testing operations on counters, and likewise pushdown automata
are able to detect empty stacks. Variants of Petri nets have also
explored emptiness testing for places, usually resulting in a great
leap in computational power. In the context of channel machines, a
slight refinement of emptiness testing is \emph{occurrence testing},
checking that a given channel contains no occurrence of a particular
message, as defined and studied in~\cite{OW06a}. Emptiness and
occurrence testing provide some measure of control over insertion
errors, since once a message has been inserted into a channel, it
remains there until it is read off it.

Our main focus in this paper is the complexity of the
\emph{termination} problem for insertion channel machines: given such
a machine, are all of its computations finite? We show that
termination is non-elementary, yet primitive recursive. This result is
quite surprising, as the closely related problems of reachability and
recurrent reachability are respectively non-primitive recursive and
undecidable. Moreover, the mere \emph{decidability} of termination for
insertion channel machines follows from the theory of well-structured
transition systems, in a manner quite similar to that for lossy
channel machines. In the latter case, however, termination is
non-primitive recursive, as shown
in~\cite{schnoebelen02}. Obtaining a primitive recursive upper
bound for insertion channel machines has therefore required us to
abandon the well-structure and pursue an entirely new approach.

On the practical side, one of the main motivations for studying
termination of insertion channel machines arises from the safety
fragment of Metric Temporal Logic. Safety MTL was shown to be
decidable in~\cite{OW06b}, although no non-trivial bounds on the
complexity could be established at the time. It is not difficult,
however, to show that (non-)termination for insertion channel machines
reduces (in polynomial time) to satisfiability for Safety MTL; the
latter, therefore, is also non-elementary. We note that in a similar
vein, a lower bound for the complexity of satisfiability of an
extension of Linear Temporal Logic was given in~\cite{Laz06}, via a
reduction from the termination problem for counter machines with
incrementation errors.

\section{Decision Problems for Faulty Channel Machines: A Brief Survey}

In this section, we briefly review some key decision problems for
lossy and insertion channel machines (the latter equipped with either
emptiness or occurrence testing). Apart from the results on
termination and structural termination for insertion channel machines,
which are presented in the following sections, all results that appear
here are either known or follow easily from known facts. Our
presentation is therefore breezy and terse. Background material on
well-structured transition systems can be found
in~\cite{FS01}.

The \emph{reachability} problem asks whether a given distinguished
control state of a channel machine is reachable. This problem was
shown to be non-primitive recursive for lossy channel machines
in~\cite{schnoebelen02}; it is likewise non-primitive recursive for
insertion channel machines via a straightforward reduction from the
latter~\cite{OW05}.

The \emph{termination} problem asks whether all computations of a
channel machine are finite, starting from the initial control state
and empty channel contents. This problem was shown to be non-primitive
recursive for lossy channel machines in~\cite{schnoebelen02}. For
insertion channel machines, we prove that termination is
non-elementary in Section~\ref{sec:non-elem} and primitive recursive
in Section~\ref{sec:prim-rec}.

The \emph{structural termination} problem asks whether all
computations of a channel machine are finite, starting from the
initial control state but regardless of the initial channel
contents. This problem was shown to be undecidable for lossy channel
machines in~\cite{MayrTCS}. For insertion channel machines, it is easy
to see that termination and structural termination coincide, so that
the latter is also non-elementary primitive-recursive decidable.

Given a channel machine $\mathcal{S}$ and two distinguished control
states $p$ and $q$ of $\mathcal{S}$, a \emph{response} property is an
assertion that every $p$ state is always eventually followed by a $q$
state in any infinite computation of $\mathcal{S}$. Note that a
counterexample to a response property is a computation that eventually
visits $p$ and forever avoids $q$ afterwards. The undecidability of
response properties for lossy channel machines follows easily from
that of structural termination, as the reader may wish to verify.

In the case of insertion channel machines, response properties are
decidable, albeit at non-primitive recursive cost (by reduction from
reachability). For decidability one first shows using the theory of
well-structured transition systems that the set of all reachable
configurations, the set of $p$-configurations, and the set of
configurations that have infinite $q$-avoiding computations are all
effectively computable. It then suffices to check whether their mutual
intersection is empty.

The \emph{recurrence} problem asks, given a channel machine and a
distinguished control state, whether the machine has a computation
that visits the distinguished state infinitely often. It is
undecidable for lossy channel machines by reduction from response, and
was shown to be undecidable for insertion channel machines
in~\cite{OW06a}.

Finally, \emph{CTL and LTL model checking} for both lossy and
insertion channel machines are undecidable, which can be established
along the same lines as the undecidability of recurrence.

These results are summarised in Figure~\ref{fig:summary_results}.

\begin{figure}
\begin{center}
\begin{tabular}{
!{\vrule width 0.7pt}c!{\vrule width 0.7pt}%
c!{\vrule width 0.7pt}c!{\vrule width 0.7pt}} \hline \cline{1-3}
& 
\textbf{Lossy Channel Machines} & \textbf{Insertion Channel Machines} \\ 
\hline \cline{1-3}
Reachability & non-primitive recursive
&non-primitive recursive\\ \hline
Termination & non-primitive recursive
&non-elementary / primitive recursive\\ \hline
Struct.\ term. & undecidable
&non-elementary / primitive recursive\\ \hline
Response & undecidable
&non-primitive recursive\\ \hline
Recurrence & undecidable
&undecidable\\ \hline
CTL / LTL & undecidable
&undecidable\\
\hline \cline{1-3}
\end{tabular}
\end{center}
\caption{Complexity of decision problems for faulty channel machines.}
\label{fig:summary_results}
\end{figure}

\section{Definitions}

A \emph{channel machine} is a tuple $\mathcal{S} =
(Q,\mathit{init},\Sigma,C,\Delta)$, where $Q$ is a finite set of
control states, $\mathit{init} \in Q$ is the initial control state,
$\Sigma$ is a finite channel alphabet, $C$ is a finite set of channel
names, and $\Delta \subseteq Q \times L \times Q$ is the transition
relation, where $L = \{ c!a, c?a, c\mbox{$=$}\emptyset,
a\mbox{$\notin$}c : c \in C, a \in \Sigma\}$ is the set of
transition labels. Intuitively, label $c!a$ denotes the writing of
message $a$ to tail of channel $c$, label $c?a$ denotes the reading of
message $a$ from the head of channel $c$, label $c\mbox{$=$}\emptyset$
tests channel $c$ for emptiness, and label $a\mbox{$\notin$}c$ tests
channel $c$ for the absence (non-occurrence) of message $a$.

We first define an \emph{error-free} operational semantics for channel
machines. Given $\mathcal{S}$ as above, a \emph{configuration} of
$\mathcal{S}$ is a pair $(q,U)$, where $q \in Q$ is the control state
and $U \in (\Sigma^*)^C$ gives the contents of each channel. Let us
write $\mathit{Conf}$ for the set of possible
configurations of $\mathcal{S}$. The rules in $\Delta$ induce an
$L$-labelled transition relation on $\mathit{Conf}$, as
follows:
\begin{itemize}
\item[(1)] $(q,c!a,q') \in \Delta$ yields a transition $(q,U)
  \progress{c!a} (q',U')$, where $U'(c) = U(c)\mbox{$\cdot$} a$ and
  $U'(d) = U(d)$ for $d \neq c$.
\emph{In other words, the channel machine moves from control state $q$
to control state $q'$, writing message $a$ to the tail of channel $c$
and leaving all other channels unchanged.}

\item[(2)] $(q,c?a,q') \in \Delta$ yields a transition $(q,U)
  \progress{c?a} (q',U')$, where $U(c) = a\mbox{$\cdot$}U'(c)$ and
  $U'(d) = U(d)$ for $d \neq c$.
\emph{In other words, the channel machine reads message $a$ from the
head of channel $c$ while moving from control state $q$ to control
state $q'$, leaving all other channels unchanged.}

\item[(3)] $(q,c\mbox{$=$}\emptyset,q') \in \Delta$ yields a transition
  $(q,U) \progress{c=\emptyset} (q',U)$, provided $U(c)$ is
the empty word.
\emph{In other words, the transition is only enabled if channel $c$ is
empty; all channel contents remain the same.}

\item[(4)] $(q,a\mbox{$\notin$}c,q') \in \Delta$ yields a transition
  $(q,U) \progress{a \notin c} (q',U)$, provided
$a$ does not occur in $U(c)$.
\emph{In other words, the transition is only enabled if channel $c$ contains
no occurrence of message $a$; all channels remain unchanged.}

\end{itemize}

If the only transitions allowed are those listed above, then we call
$\mathcal{S}$ an \emph{error-free} channel machine. This machine model
is easily seen to be Turing powerful~\cite{BZ83}. As discussed
earlier, however, we are interested in channel machines with
(potential) \emph{insertion errors}; intuitively, such errors are
modelled by postulating that channels may at any time acquire
additional messages interspersed throughout their current contents.

For our purposes, it is convenient to adopt the \emph{lazy} model of
insertion errors, given next. Slightly different models, such as
those of \cite{CFP96,OW06a}, have also appeared in the
literature. As the reader may easily check, all these models are
equivalent insofar as reachability and termination properties are
concerned.

The lazy operational semantics for channel machines with insertion
errors simply augments the transition relation on $\mathit{Conf}$ with
the following rule:
\begin{itemize}
\item[(5)] $(q,c?a,q') \in \Delta$ yields a transition $(q,U)
  \progress{c?a} (q',U)$. 
\emph{In other words, insertion errors occur `just in time',
  immediately prior to a read operation; all channel contents remain
  unchanged.}
\end{itemize}

The channel machines defined above are called \emph{insertion channel
machines with occurrence testing}, or \emph{ICMOT}s.
We will also consider \emph{insertion channel machines with emptiness
testing}, or \emph{ICMET}s. The latter are simply ICMOTs without any
occurrence-testing transitions (i.e., transitions labelled with
$a\mbox{$\notin$}c$).

A \emph{run} of an insertion channel machine is a finite or infinite
sequence of transitions of the form $\sigma_0 \progress{l_0} \sigma_1
\progress{l_1} \ldots$ that is consistent with the lazy operational
semantics. The run is said to start from the initial configuration if
the first control state is $\mathit{init}$ and all channels are
initially empty.

Our main focus in this paper is the study of the complexity of the
\emph{termination} problem: given an insertion channel machine
$\mathcal{S}$, are all runs of $\mathcal{S}$ starting from the initial
configuration finite?

\section{Termination is Non-Elementary}
\label{sec:non-elem}

In this section, we show that the termination problem for insertion
channel machines---ICMETs and ICMOTs---is non-elementary. More
precisely, we show that the termination problem for ICMETs of size $n$
in the worst case requires time at least
$2\mbox{$\Uparrow$}\Omega(\log n)$.\footnote{The expression
$2\mbox{$\Uparrow$}m$,
known as 
tetration, denotes an exponential tower of $2$s of height $m$.} 
Note that the same immediately follows for ICMOTs.

Our proof proceeds by reduction from the termination problem for
two-counter machines in which the counters are tetrationally bounded;
the result then follows from standard facts in complexity theory (see,
e.g.,~\cite{HU79}).

Without insertion errors, it is clear that a channel machine can
directly simulate a two-counter machine simply by storing the values
of the counters on one of its channels. To simulate a counter machine
in the presence of insertion errors, however, we require periodic
integrity checks to ensure that the representation of the counter
values has not been corrupted.  Below we give a simulation that
follows the `yardstick' construction of Meyer and
Stockmeyer~\cite{SM73,LNORW07}: roughly speaking, we use an
$m$-bounded counter to check the integrity of a $2^m$-bounded counter.

\begin{theorem}
\label{lower_bound_theorem}
The termination problem for ICMETs and ICMOTs is non-elementary. 
\end{theorem}

\proof Let us say that a counter is $m$-bounded if it can take values
in $\{0,1,\ldots,m-1\}$.  We assume that such a counter $u$ comes
equipped with procedures $\textsc{Inc}(u)$, $\textsc{Dec}(u)$,
$\textsc{Reset}(u)$, and $\textsc{IsZero}(u)$, where
$\textsc{Inc}$ and $\textsc{Dec}$ operate modulo $m$, and increment,
resp.\ decrement, the counter.  We show how to simulate a
deterministic counter machine $\mathcal{M}$ of size $n$ equipped with
two $2 \mbox{$\Uparrow$} n$-bounded counters by an ICMET
$\mathcal{S}$ of size $2^{O(n)}$.  We use this simulation to reduce the
termination problem for $\mathcal{M}$ to the termination problem for
$\mathcal{S}$.

By induction, assume that we have constructed an ICMET $\mathcal{S}_k$
that can simulate the operations of a $2 \mbox{$\Uparrow$} k$-bounded
counter $u_k$.  We assume that $\mathcal{S}_k$ correctly implements
the operations $\textsc{Inc}(u_k)$, $\textsc{Dec}(u_k)$,
$\textsc{Reset}(u_k)$, and $\textsc{IsZero}(u_k)$ (in particular,
we assume that the simulation of these operations by $\mathcal{S}_k$
is guaranteed to terminate). We describe an ICMET $\mathcal{S}_{k+1}$
that implements a $2 \mbox{$\Uparrow$} (k+1)$-bounded counter
$u_{k+1}$.  $\mathcal{S}_{k+1}$ incorporates $\mathcal{S}_k$, and thus
can use the above-mentioned operations on the counter $u_k$ as
subroutines.  In addition, $\mathcal{S}_{k+1}$ has two extra channels
$c$ and $d$ on which the value of counter $u_{k+1}$ is stored in
binary.  We give a high-level description.

We say that a configuration of $\mathcal{S}_{k+1}$ is \emph{clean} if
channel $c$ has size $2 \mbox{$\Uparrow$} k$ and channel $d$ is empty.
We ensure that all procedures on counter $u_{k+1}$ operate correctly
when they are invoked in clean configurations of $\mathcal{S}_{k+1}$, and
that they also yield clean configurations upon completion.  In fact,
we only give details for the procedure $\textsc{Inc}(u_{k+1})$---see
Figure~\ref{fig:inc}; the others should be clear from this example.

\begin{figure}
\textbf{Procedure} $\textsc{Inc}(u_{k+1})$\hspace{\fill}

\ \ \ $\textsc{Reset}(u_k)$\hspace{\fill}

\ \ \ \textbf{repeat}\hspace{\fill}

\ \ \ \ \ \ $c?x \, ; \, d!(1-x)$\ \ \ \ \ /* Increment counter
$u_{k+1}$ while transferring $c$ to $d$ */\hspace{\fill}\hspace{\fill}

\ \ \ \ \ \ $\textsc{Inc}(u_k)$\hspace{\fill}

\ \ \ \textbf{until} $\textsc{IsZero}(u_k)$ or $x=0$\hspace{\fill}
      
\ \ \ \textbf{while} not $\textsc{IsZero}(u_k)$
\textbf{do}\hspace{\fill}

\ \ \ \ \ \ $c?x \, ; \, d!x$\ \ \ \ \ \ \ \ \ \ \ \,
                             /* Transfer remainder of $c$ to $d$
                                                      */\hspace{\fill}

\ \ \ \ \ \ $\textsc{Inc}(u_k)$\hspace{\fill}

\ \ \ \textbf{endwhile}\hspace{\fill}

\ \ \ $\mathbf{test}(c\mbox{$=$}\emptyset)$\ \ \ 
\ \ \ \ \ \ \ \ \ \ /* Check that there were no
insertion errors on $c$, otherwise halt */ \hspace{\fill}

\ \ \ \textbf{repeat}\hspace{\fill}

\ \ \ \ \ \ $d?x \, ; \, c!x$\ \ \ \ \ 
\ \ \ \ \ \ \ /* Transfer $d$ back to $c$ */ \hspace{\fill}

\ \ \ \ \ \ $\textsc{Inc}(u_k)$\hspace{\fill}

\ \ \ \textbf{until} $\textsc{IsZero}(u_k)$\hspace{\fill}

\ \ \ $\mathbf{test}(d\mbox{$=$}\emptyset)$\ \ \ \ 
\ \ \ \ \ \ \ \, /* Check that there were no
insertion errors on $d$, otherwise halt */ \hspace{\fill}

\ \ \ $\mathbf{return}$\hspace{\fill}

\caption{Procedure to increment counter $u_{k+1}$. Initially, this
  procedure assumes that counter $u_{k+1}$ is encoded in binary on
  channel $c$, with least significant bit at the head of the channel;
  moreover, $c$ is assumed to comprise exactly $2 \mbox{$\Uparrow$} k$
  bits (using padding 0s if need be). In addition, channel $d$ is
  assumed to be initially empty. Upon exiting, channel $c$ will
  contain the incremented value of counter $u_{k+1}$ (modulo $2
  \mbox{$\Uparrow$} (k+1)$) in binary, again using $2
  \mbox{$\Uparrow$} k$ bits, and channel $d$ will be empty. We
  regularly check that no insertion errors have occurred on channels
  $c$ or $d$ by making sure that they contain precisely the right
  number of bits. This is achieved using counter $u_{k}$ (which can
  count up to $2 \mbox{$\Uparrow$} k$ and is assumed to work
  correctly) together with emptiness tests on $c$ and $d$. If an
  insertion error does occur during execution, the procedure will
  either halt, or the next procedure to handle channels $c$ and $d$
  (i.e., any command related to counter $u_{k+1}$) will halt.}

\label{fig:inc}
\end{figure}

Since the counter $u_k$ is assumed to work correctly, the above
procedure is guaranteed to terminate, having produced the correct
result, in the absence of any insertion errors on channels $c$ or
$d$. On the other hand, insertion errors on either of these channels
will be detected by one of the two emptiness tests, either immediately
or in the next procedure to act on them.

The initialisation of the induction is handled using an ICMET
$\mathcal{S}_1$ with no channel (in other words, a finite automaton)
of size $2$, which can simulate a $2$-bounded counter (i.e., a single
bit). The finite control of the counter machine, likewise, is duplicated
using a further channel-less ICMET\@.

Using a product construction, it is straightforward to conflate these
various ICMETs into a single one, $\mathcal{S}$, of size exponential
in $n$ (more precisely: of size $2^{O(n)}$). As the reader can easily
check, $\mathcal{M}$ has an infinite computation iff $\mathcal{S}$
has an infinite run. The result follows immediately. \qed

\section{Termination is Primitive Recursive}
\label{sec:prim-rec}

The central result of our paper is the following:

\begin{theorem}
\label{upper_bound_theorem}
The termination problem for ICMOTs and ICMETs is primitive
recursive. More precisely, when restricting to the class of ICMOTs or
ICMETs that have at most $k$ channels, the termination problem is in
$(k\mbox{$+$}1)$-EXPSPACE.
\end{theorem}

\proof

In what follows, we sketch the proof for ICMOTs, ICMETs being a
special case of ICMOTs. Let us also assume that our ICMOTs do not make
use of any emptiness tests; this restriction is harmless since any
emptiness test can always be replaced by a sequence of occurrence
tests, one for each letter of the alphabet, while preserving
termination.

Let $\mathcal{S} = (Q,\mathit{init},\Sigma,C,\Delta)$ be a fixed ICMOT
without emptiness tests; in other words, $\mathcal{S}$'s set of
transition labels is $L = \{ c!a, c?a, a\mbox{$\notin$}c : c \in C, a
\in \Sigma\}$. Our strategy is as follows: we suppose that
$\mathcal{S}$ has no infinite runs, and then derive an upper bound on
the length of the longest possible finite run. The result follows by
noting that the total number of possible runs is exponentially bounded
by this maximal length.

For a subset $D \subseteq C$ of channels, we define an equivalence
$\equiv_D$ over the set $\mathit{Conf}$ of configurations of
$\mathcal{S}$ as follows:
\[
  (q,U) \equiv_D (q',U')\ \ \text{iff}\ \ q=q'\ \text{and}\
  U(d)=U'(d)\ \text{for every}\ d \in D.
\] 

Let us write $\mathit{Conf}_D$ to denote the set
$\mathit{Conf}/\mbox{$\equiv$}_D$ of equivalence classes of
$\mathit{Conf}$ with respect to $\equiv_D$. Furthermore, given $f : D
\rightarrow \mathbb{N}$ a `bounding function' for the channels in $D$,
let
\[ \mathit{Conf}_D^f = \{[(q,U)]_D \in \mathit{Conf}_D
: |U(d)| \leq f(d)\ \text{for every}\ d \in D\}
\] 
be the subset of $\mathit{Conf}_D$ consisting of those equivalence
classes of configurations whose $D$-channels are bounded by $f$. As
the reader can easily verify, we have the following bound on the
cardinality $\gamma_D^f$ of $\mathit{Conf}_D^f$:
\begin{equation}
\label{gamma-bound}
\gamma_D^f \leq |Q| \prod_{d \in D}(|\Sigma|+1)^{f(d)}.
\end{equation} 

Consider a finite run $\sigma_0 \progress{l_0} \sigma_1 \progress{l_1}
\ldots \progress{l_{n-1}} \sigma_n$ of $\mathcal{S}$ (with $n \geq
1$), where each $\sigma_i \in \mathit{Conf}$ is a configuration and
each $l_i \in L$ is a transition label. We will occasionally write
$\sigma_0 \Progress{\lambda} \sigma_n$ to denote such a run, where
$\lambda = l_0 l_1 \ldots l_{n-1} \in L^+$.

We first state a pumping lemma of sorts, whose straightforward proof
is left to the reader:

\begin{lemma}
\label{lem-iteration}
Let $D \subseteq C$ 
be given, and assume that $\sigma\Progress{\lambda}\sigma'$ (with
$\lambda \in L^+$) is a run of $\mathcal{S}$ such that
$\sigma\equiv_D\sigma'$. Suppose further that, for every label $a
\mbox{$\notin$} c$ occurring in $\lambda$, either $c \in D$, or the
label $c!a$ does not occur in $\lambda$. Then $\lambda$ is repeatedly
firable from $\sigma$, i.e., there exists an infinite run $\sigma
\Progress{\lambda}\sigma'\Progress{\lambda}\sigma''\Progress{\lambda}\ldots$.
\end{lemma}

Note that the validity of Lemma~\ref{lem-iteration} rests crucially on
(the potential for) insertion errors.

Let $\langle w_i\rangle_{1 \leq i \leq n}$ be a finite sequence, and
let $0 < \alpha \leq 1$ be a real number. A set $S$ is said to
be $\alpha$-frequent in the sequence $\langle w_i \rangle$ if the set
$\{i : w_i \in S\}$ has cardinality at least $\alpha n$. 

The next result we need is a technical lemma guaranteeing a certain
density of repeated elements in an $\alpha$-frequent sequence:

\begin{lemma}
\label{lem-density}
Let $\langle w_i\rangle_{1 \leq i \leq n}$ be a finite sequence, and
assume that $S$ is a finite $\alpha$-frequent set in $\langle w_i
\rangle$. Then there exists a sequence of pairs of indices $\langle
(i_j, i'_j) \rangle_{1 \leq j \leq \frac{\alpha n}{2(|S|+1)}}$ 
such that, for all $j < \frac{\alpha n}{2(|S|+1)}$, we have $i_j<i'_j <
i_{j+1}$, $i'_j-i_j \leq \frac{2(|S|+1)}{\alpha}$, and $w_{i_j} =
w_{i'_j} \in S$.
\end{lemma}

\proof By assumption, $\langle w_i\rangle$ has a subsequence of length
at least $\alpha n$ consisting exclusively of elements of $S$. This
subsequence, in turn, contains at least $\frac{\alpha n}{|S|+1}$
disjoint `blocks' of length $|S|+1$. By the pigeonhole principle, each
of these blocks contains at least two identical elements from $S$,
yielding a sequence of pairs of indices $\langle (i_j, i'_j)
\rangle_{1 \leq j \leq \frac{\alpha n}{|S|+1}}$ having all the
required properties apart, possibly, from the requirement that $i'_j -
i_j \leq \frac{2(|S|+1)}{\alpha}$. Note also that there are, for now,
twice as many pairs as required.

Consider therefore the half of those pairs whose difference is
smallest, and let $p$ be the largest such difference. Since the other
half of pairs in the sequence $\langle (i_j, i'_j) \rangle$ have
difference at least $p$, and since there is no overlap between
indices, we have $\frac{1}{2}\cdot \frac{\alpha n}{|S|+1}\cdot p < n$,
from which we immediately derive that $p$ is bounded by
$\frac{2(|S|+1)}{\alpha}$, as required.
This concludes the proof of Lemma~\ref{lem-density}. \qed

Recall our assumption that $\mathcal{S}$ has no infinite run, and let
$\pi = \sigma_0 \progress{l_0} \sigma_1 \progress{l_1} \ldots
\progress{l_{n-1}} \sigma_n$ be any finite run of $\mathcal{S}$,
starting from the initial configuration; we seek to obtain an upper
bound on $n$.

Given a set $D \subseteq C$ of channels, it will be convenient to
consider the sequence $[\pi]_D = \langle [\sigma_i]_D \rangle_{0 \leq
i \leq n}$ of equivalence classes of configurations in $\pi$ modulo
$\equiv_D$ (ignoring the interspersed labelled transitions for now).

Let $f : C \rightarrow \mathbb{N}$ and $0 < \alpha \leq 1$ be given,
and suppose that $\mathit{Conf}_C^f$ is $\alpha$-frequent in
$[\pi]_C$, so that there are at least $\alpha n$ occurrences of
configuration equivalence classes in $\mathit{Conf}_C^f$ along
$[\pi]_C$. Recall that $\mathit{Conf}_C^f$ contains $\gamma_C^f$
elements. Observe, by Lemma~\ref{lem-iteration}, that no member of
$\mathit{Conf}_C^f$ can occur twice along $[\pi]_D$, otherwise
$\mathcal{S}$ would have an infinite run.  Consequently,
\begin{equation}
\label{n-bound}
n \leq \frac{\gamma_C^f}{\alpha}.
\end{equation}

We will now inductively build an increasing sequence 
$\emptyset = D_0 \subset D_1 \subset \ldots \subset D_{|C|} = C$, as well
as functions $f_i : D_i \rightarrow \mathbb{N}$ and real numbers $0 <
\alpha_i \leq 1$, for $0 \leq i \leq |C|$, such that
$\mathit{Conf}_{D_i}^{f_i}$ is $\alpha_i$-frequent in $[\pi]_{D_i}$ for
every $i \leq |C|$. 

The base case is straightforward: the set
$\mathit{Conf}_\emptyset^{f_0} = \mathit{Conf}_\emptyset$ is clearly
$1$-frequent in $[\pi]_\emptyset$.

Let us therefore assume that $\mathit{Conf}_D^f$ is $\alpha$-frequent
in $[\pi]_D$ for some strict subset $D$ of $C$ and some $f : D
\rightarrow \mathbb{N}$ and $\alpha > 0$. We now compute $D' \subseteq C$
strictly containing $D$, $f' : D' \rightarrow \mathbb{N}$, and
$\alpha' > 0$ such that $\mathit{Conf}_{D'}^{f'}$ is
$\alpha'$-frequent in $[\pi]_{D'}$.

Thanks to our induction hypothesis and Lemma~\ref{lem-density}, we
obtain a sequence of pairs of configurations $\langle
(\theta_j,\theta'_j)\rangle_{1 \leq j \leq h}$, where $h =
\frac{\alpha n}{2(\gamma_D^f+1)}$, $[\theta_j]_D = [\theta'_j]_D \in
\mathit{Conf}_D^f$, and such that
\[\pi = \sigma_0 \Longrightarrow \theta_1 \Progress{\lambda_1} \theta'_1 
\Longrightarrow \theta_2 \Progress{\lambda_2} \theta'_2
\Longrightarrow \ldots \Longrightarrow \theta_h \Progress{\lambda_h} 
\theta'_h \Longrightarrow \sigma_n
\]
with each $\lambda_j \in L^+$ having length no greater than 
$\frac{2(\gamma_D^f + 1)}{\alpha}$, for $1 \leq j \leq h$.

For each $\lambda_j$, let $\mathit{OT}_j$ be the set of
occurrence-test labels that occur at least once in $\lambda_j$. Among
these sets, let $\mathit{OT}$ denote the one that appears most
often. Note that there are $2^{|\Sigma|\cdot|C|}$ different possible
sets of occurrence-test labels, and therefore at least
$\frac{h}{2^{|\Sigma|\cdot|C|}}$ of the $\mathit{OT}_j$ are equal to 
$\mathit{OT}$. 

Following a line of reasoning entirely similar to that used in
Lemma~\ref{lem-density}\footnote{Formally, we could directly invoke
Lemma~\ref{lem-density}, as follows. Write the sequence of transition
labels of $\pi$ as
$\delta_0\lambda_1\delta_1\lambda_2\cdots\lambda_h\delta_h$, with
the $\lambda_i$ as above. Next, formally replace each instance of
$\lambda_i$ whose set of occurrence-test labels is $\mathit{OT}$ by a
new symbol $\mathbf{O}$; if needed, add dummy non-$\mathbf{O}$ symbols
to the end of the sequence to bring its length up to $n$, and call the
resulting sequence $\langle w_i \rangle$. Finally, note that the
singleton set $\{\mathbf{O}\}$ is
$\frac{h}{2^{|\Sigma|\cdot|C|}\cdot n}$-frequent in $\langle w_i
\rangle$.}, we can deduce that $\pi$ contains at least
$\frac{h}{4\cdot2^{|\Sigma|\cdot|C|}} = \frac{\alpha n}
{8(\gamma_D^f+1)2^{|\Sigma|\cdot|C|}}$ non-overlapping patterns of the
form
\[ \theta \Progress{\lambda} \theta' 
\Progress{\delta} \bar{\theta} \Progress{\bar{\lambda}} \bar{\theta}', \]
where:
\begin{itemize}

\item $[\theta]_D = [\theta']_D \in \mathit{Conf}_D^f$
and $[\bar{\theta}]_D = [\bar{\theta}']_D \in \mathit{Conf}_D^f$,

\item $\lambda,\bar{\lambda} \in L^+$ each have length no greater
than $\frac{2(\gamma_D^f + 1)}{\alpha}$,

\item $\delta \in L^+$ has length no greater than 
$\frac{8(\gamma_D^f+1)2^{|\Sigma|\cdot|C|}}{\alpha}$, and

\item the set of occurrence-test labels occurring in $\lambda$ and
$\bar{\lambda}$ in both cases is $\mathit{OT}$.

\end{itemize}

Consider such a pattern. Observe that $\lambda$ must contain at least
one occurrence-test label $a \mbox{$\notin$} c$ with $c \notin D$ and
such that the label $c!a$ occurs in $\lambda$, otherwise $\mathcal{S}$
would have an infinite run according to
Lemma~\ref{lem-iteration}. Pick any such occurrence-test label and let
us denote it $a \mbox{$\notin$} c$.

We now aim to bound the size of channel $c$ in the $\bar{\theta}$
configuration of our patterns. Note that since $\lambda$ and
$\bar{\lambda}$ contain the same set of occurrence-test labels, the
label $a \mbox{$\notin$} c$ occurs in $\bar{\lambda}$. That is to say,
somewhere between configurations $\bar{\theta}$ and $\bar{\theta}'$,
we know that channel $c$ did not contain any occurrence of $a$. On the
other hand, an $a$ was written to the tail of channel $c$ at some
point between configurations $\theta$ and $\theta'$, since $\lambda$
contains the label $c!a$. For that $a$ to be subsequently read off the
channel, the whole contents of channel $c$ must have been read from
the time of the $c!a$ transition in $\lambda$ to the time of the $a
\mbox{$\notin$} c$ transition in $\bar{\lambda}$. Finally, note that,
according to our lazy operational semantics, the size of a channel
changes by at most 1 with each transition. It follows that the size of
channel $c$ in configuration $\bar{\theta}$ is at most $|\lambda| +
|\delta| + |\lambda'| \leq \frac{(\gamma_D^f+1)(4+8\cdot
2^{|\Sigma|\cdot|C|})}{\alpha}$.

Let $D' = D \cup \{c\}$, and define the bounding function $f':D'
\rightarrow \mathbb{N}$ such that $f'(d) = f(d)$ for all $d \in D$,
and $f'(c) = \frac{(\gamma_D^f+1)(4+8\cdot
2^{|\Sigma|\cdot|C|})}{\alpha}$. From our lower bound on the number of
special patterns, we conclude that the set $\mathit{Conf}_{D'}^{f'}$
is $\alpha'$-frequent in $[\pi]_{D'}$, where $\alpha' = \frac{\alpha}
{8(\gamma_D^f+1)2^{|\Sigma|\cdot|C|}}$.

We now string everything together to obtain a bound on $n$, the length
of our original arbitrary run $\pi$. For convenience, let
$c_1,c_2,\ldots,c_{|C|}$ be an enumeration of the channel names in $C$
in the order in which they are picked in the course of our proof; thus
$D_i = D_{i-1} \cup \{c_i\}$ for $1 \leq i \leq |C|$. Correspondingly,
let $M_i = f_{i}(c_i)$, for $0 \leq i \leq |C|$, with the convention
that $M_0 = 1$; it is easy to see that $M_i$ is the maximum value of
$f_i$ over $D_i$, since the sequences $\langle
\gamma_{D_i}^{f_i}\rangle$ and $\langle \alpha_i \rangle$ are
monotonically increasing and decreasing respectively. 

From Equation~\ref{gamma-bound}, we easily get that
$\gamma_{D_i}^{f_i} \in O(|\mathcal{S}|^{|\mathcal{S}|M_i})$, where
$|\mathcal{S}|$ is any reasonable measure of the size of our ICMOT
$\mathcal{S}$. Combining this with our expressions for $f'$ and
$\alpha'$ above, we obtain that
$ 
M_{i+1}, \frac{1}{\alpha_{i+1}} \in 
O\left(\frac{|\mathcal{S}|^{|\mathcal{S}|^2 M_i}}{\alpha_i}\right)
$ 
for $0 \leq i \leq |C|-1$. This, in turns, lets us derive bounds for 
$\gamma_{C}^{f_{|C|}}$ and $\alpha_{|C|}$, which imply, together with
Equation~\ref{n-bound}, that
\[ n \leq 2^{2^{\cdot^{\cdot^{\cdot^{2^{P(|\mathcal{S}|)}}}}}},\]
where $P$ is some polynomial (independent of $\mathcal{S}$), and 
the total height of the tower of exponentials is $|C| + 2$.

The ICMOT $\mathcal{S}$ therefore has an infinite run iff it has a run
whose length exceeds the above bound. Since the lazy operational
semantics is finitely branching (bounded, in fact, by the size of the
transition relation), this can clearly be determined in
$(|C|\mbox{$+$}1)$-EXPSPACE, which concludes the proof of
Theorem~\ref{upper_bound_theorem}. \qed

Theorems~\ref{lower_bound_theorem} and \ref{upper_bound_theorem}
immediately entail the following:

\begin{corollary}
The structural termination problem---are all computations of the 
machine finite, starting from the initial control state but
regardless of the initial channel contents?---is decidable for ICMETs
and ICMOTs, with non-elementary but primitive-recursive complexity.
\end{corollary}

\section{Conclusion}

The main result of this paper is that termination for
insertion channel machines with emptiness or occurrence testing has
non-elementary, yet primitive recursive complexity. This result is in
sharp contrast with the equivalent problem for lossy channel machines,
which has non-primitive recursive complexity.

We remark that the set of configurations from which a given insertion
channel machine has at least one infinite computation is finitely
representable (thanks to the theory of well-structured transition
systems), and is in fact computable as the greatest fixed point of the
pre-image operator. The proof of Theorem~\ref{upper_bound_theorem},
moreover, shows that this fixed point will be reached in
primitive-recursively many steps. The set of configurations from which
there is an infinite computation is therefore primitive-recursively
computable, in contrast with lossy channel machines for which it is
not even recursive (as can be seen from the undecidability of
structural termination).

Finally, another interesting difference with lossy channel machines
can be highlighted by quoting a slogan from~\cite{schnoebelen02}:
``\emph{Lossy systems with $k$ channels can be [polynomially] encoded
into lossy systems with one channel.}'' We can deduce from
Theorems~\ref{lower_bound_theorem} and \ref{upper_bound_theorem} that
any such encoding, in the case of insertion channels machines, would
require non-elementary resources to compute, if it were to preserve
termination properties.

\end{document}